 \documentclass{camera}

\usepackage{graphicx}

\newcount\flag
\input epsf

\begin{document}


\title{THE ORIGIN OF COSMIC RAYS -\\
 A 96-YEAR-OLD PUZZLE SOLVED?}

%

\author{Arnon Dar}


\organization{Department of Physics and Space Research 
Institute,\\Technion, Haifa 32000, Israel}

\maketitle

\begin{abstract}

There is mounting evidence that long duration gamma ray bursts (GRBs) are
produced by ultra-relativistic jets of ordinary matter which are ejected
in core collapse supernova (SN) explosions.  Such jets are extremely
efficient cosmic ray (CR) accelerators which can accelerate the swept up
ambient particles on their way to the highest observed CR energies.  The
bulk of the jet kinetic energy is used to accelerate CRs while only a tiny
fraction is used to produce the GRB and its afterglow.  Here we use the
remarkably successful cannonball (CB) model of GRBs to show that the
bipolar jets from SN explosions, which produce GRBs most of which are not
beamed towards Earth, can be the main origin of cosmic rays at all
energies. The model explains very simply the elemental composition of CRs
and their observed spectra at all energies. In particular it explains the
origin of the CR knees and ankle.  Above the CR ankle, the Galactic
magnetic fields can no longer delay the free escape of ultra-high energies
CRs (UHECR) from the Galaxy, and the CRs from the intergalactic medium
(IGM), which were injected there by SN jets from all the galaxies and
isotropized there by the IGM magnetic fields, dominate the Galactic CR
spectrum.  A Greisen-Zatsepin-Kuzmin (GZK) cutoff due to the interaction
of UHECRs with the microwave background radiation is expected. The CR
nuclei which diffuse out of galaxies, or are directly deposited in the IGM
by the relativistic SN jets, may be the origin of the IGM magnetic fields. 
Inverse Compton scattering of the cosmic microwave background radiation
(MBR) by the CR electrons in the IGM produces the diffuse extragalactic
gamma-ray background radiation (GBR).

\end{abstract}
\section{Introduction}
\noindent
Cosmic rays (CRs) were discovered by Victor Hess in 1912. Today, 92 years
later, their origin is still unknown.  CRs have been studied in
experiments above the atmosphere, in the atmosphere, on the ground,
underground and in space. Their energies cover an enormous range, from sub
GeV to more than a few $10^{11}\, GeV\,,$ over which their differential
flux decreases by roughly 33 orders of magnitude. Any successful theory of
the origin of CRs must explain their main observed properties near Earth
(for recent reviews see, e.g. 
Biermann \& Sigl 2001; Watson 2001,2003; Olinto 2004; Cronin 2004) --

\noindent
{\bf The CR energy spectrum} shown in Fig.1 has
been measured up to hundreds of $EeV$. It can be
approximated by a broken power-law, $dn/dE\sim E^{-p}\,,$ with a series of
breaks near $ 3\, PeV$ known as `the knee', a second `knee' near
200 PeV and an `ankle' near $4\, EeV\, $. The power-law index
changes from $p\sim 2.67\, $ below the knee to $p\sim 3.05$ above it and
steepens to $p\sim 3.2$ at the second knee. At the ankle, the
spectral index changes to $p\sim 2.7\, .$ The spectral behaviour
above $50\ EeV$ is still debated (see, e.g. Olinto 2004; Cronin 2004).

\noindent 
{\bf The CR elemental composition} is known well only from direct
measurements on board satellites, which run out of statistics well below
the knee energy. The measured composition of high energy CRs is highly
enriched in elements heavier than hydrogen relative to that of the solar
system, as shown in Table I for TeV CRs.  The enrichment increases with
atomic number and with CR energy almost up to the second knee, beyond
which it appears to decline (e.g., Kampert et al.~2004; Hoerandel 2004 ).
It is still debated at energies above the ankle (e.g. Cronin 2004).  The
detailed elemental composition at the knee and above it is known only very
roughly (e.g. Hoerandel 2004; Watson 2001, 2003; Cronin 2004).

\noindent
{\bf The CR arrival directions} at energy below $\sim EeV$ are 
isotropized by the Galactic magnetic fields. At energies
well above $EeV\,,$ their arrival directions 
may point towards their Galactic sources and at extremely high 
energies they may point at extragalactic sources.  
Initial reports of deviations from isotropy in the arrival directions 
of such cosmic rays with energy above $EeV$ and of some clustering in their 
arrival directions are still debated (e.g. Watson 2001,2003; Cronin 2004).

\noindent
{\bf The Galactic cosmic ray luminosity} exceeds 
$10^{42}\, erg\, s^{-1}$ (Dar \& De R\'ujula 2001b).  

\bigskip

\noindent
It is widely believed that Galactic CRs with energy below the knee are
accelerated mainly in Galactic supernova remnants (SNRs). The opinions on
the origin of CRs with energy between the knee and the ankle are still
divided between Galactic and extragalactic origin.  CRs with energy above
the ankle are generally believed to be extragalactic in origin because
they can no longer be isotropized by the Galactic magnetic fields while
their arrival directions are isotropic to a fair approximation. Yet
Galactic origin is not ruled out -- they may be produced by the decay of
unknown massive particles or other unknown sources which are distributed
isotropically in an extended Galactic halo.  An observational proof that
the CRs above the ankle are extragalactic in origin, such as arrival
directions which are correlated with identified extragalactic sources or a
Greisen-Zatsepin-Kuzmin (GZK) cutoff (Greisen 1966; Zatsepin \& Kuzmin 1966)
due to $\pi$ production in their
collisions with the cosmic microwave background radiation (MBR), are still
lacking (Cronin 2004). 
Moreover, there is no single solid observational evidence which supports
the SNR origin of CRs below the knee (see, e.g. Plaga 2002 and references
therein).  In fact, the evidence from $\gamma$-ray astronomy, X-ray
astronomy and radio astronomy strongly suggests that {\bf SNRs are not the
major accelerators of CRs with energy below the knee}: 

\begin{itemize} 
\item{}
{\bf SNR origin cannot explain the Galactic CR luminosity:}
Radio emission and X-ray emission from SNRs provide strong evidence 
for acceleration of high energy electrons in SNRs. Some SNRs were
also detected in TeV $\gamma$-rays which could be produced by the prompt 
decay of $\pi^0$'s from collisions of high energy hadronic 
cosmic rays with the ambient protons and nuclei in and around the SNR. 
However, recent observations of SNRs inside molecular clouds 
(and careful analysis of Tev emission in others) 
show that the time-integrated CR luminosity of these SNRs does not   
exceed $10^{48}\, erg\, .$ The rate of SN explosions in the Galaxy which
is estimated to be $\sim 1/50\, year,$ yields a total CR
luminosity which falls short of the estimated luminosity 
of the Milky Way (MW) in CRs, $L_{CR}[MW] > 10^{41}\, erg\, s^{-1}\, ,$
by  two orders of magnitude. 
\item{} 
{\bf SNR origin cannot explain the diffuse Galactic GBR:} 
As the CRs from SNRs diffuse through the Galactic magnetic fields, the
interactions of CR electrons with ambient photons and of CR nuclei with
ambient nuclei in the interstellar medium produce a diffuse background of
$\gamma$-rays. Such a diffuse gamma background radiation (GBR) has been
detected by EGRET and Comptel on board the Compton Gamma Ray Observatory
(CGRO). However, the scale length of the distribution of SNR in the
Galactic disk is $\sim 4\, kpc$ and cannot explain the scale length of the
observed GBR, which is larger by more than an order of magnitude
(Strong \& Mattox 1996). In
particular, energetic electrons cool rapidly by inverse Compton scattering
of stellar light and of microwave background radiation (MBR). Over their
cooling time they cannot reach 
far enough from the SNRs by diffusion to
explain the intensity of GBR at large distances from the
Galactic center 

\item{}
{\bf SNR origin cannot explain the diffuse radio emission from galaxies
and clusters:} Because of their fast cooling in the MBR the electrons from
SNRs, which are mainly located in the galactic disks, cannot reach large
distances from the galactic disks, by either diffusion or non-relativistic
winds. The radio emission from our Galaxy, from edge-on galaxies and from
the intergalactic space in clusters of galaxies provide evidence for high
energy electrons at very large distances from the galactic disks where
most of the SNRs are located. 
\end{itemize}

\noindent 

{\bf Although supernova remnants (SNRs) do not seem to be the main source
of CRs in our Galaxy, in external galaxies and in the intergalactic medium
(IGM), SN explosions may still be the main source of CRs at all energies,
if SNe emit highly relativistic bi-polar jets which produce the visible
GRBs when they point in our directions} (Dar 1998a; Dar and Plaga 
1999)\footnote{The
association between GRBs and high energy CRs was first suggested by Dar et
al. (1992). Waxman (1995), Vietri (1995) and Milgrom (1995) suggested that
extragalactic GRBs accelerate the ultra-high energy cosmic rays (UHECR)
observed near Earth while Dar (1998a) and Dar \& Plaga (1999), following the 
first observational evidence for a GRB-SN association (Galama et al 1998),
suggested that the bulk of the cosmic rays {\bf at all energies} are
accelerated in bipolar jets which are ejected in SN explosions and produce
Galactic GRBs, most of which are beamed away from Earth.}. Here, I will
outline briefly a simple theory of the origin of CRs at all energies,
which is based on the extremely successful cannonball (CB) model of GRBs
(Dar \& De R\'ujula 2000, 2003). Many ideas are adopted from 
Dar \& Plaga 1999. I will show that the theory explains
remarkably well the main observed properties of Galactic and extragalactic
cosmic rays.  A more complete theory and more rigorous derivation of CR
properties from the CB model will be published elsewhere (De R\'ujula and
Dar \& De R\'ujula, in preparation). According to this theory: 

\begin{itemize} 
\item{} 

CRs with energy below the CR ankle are Galactic in origin. 
They are ISM particles which were swept up and
accelerated by the CBs to a maximal lab energy -- the elemental
knee -- mainly by elastic magnetic deflection in the CBs rest frame.  
The first knee in the all-particle spectrum is the maximal lab
energy of CR protons; the `second' knee is that of the iron group nuclei
and heavier metals \footnote{Jets from microquasars and active galactic
nuclei may also contribute significantly to CR acceleration in galaxies,
and in clusters of galaxies (e.g. Dar 1998b; Heinz \& Sunyaev 2002), 
but they will not be discussed here.}. 

\item{} 
A fraction of the CR nuclei which are injected into the ISM with energy 
below the knee are reaccelerated by CBs from other SN explosions 
all the way to the highest observed cosmic ray energies.
CR nuclei with energy above the ankle are extragalactic in origin.  They
are galactic CRs which were reaccelerated in the interstellar medium (ISM)
by CBs from SN explosions and escaped directly into the intergalactic
medium (IGM).  They have accumulated there during the Hubble time and were
isotropized by the IGM magnetic fields.  The ankle is the energy around
which magnetic trapping and isotropization of Galactic CRs
cease to be efficient, the CRs begin to escape freely into the IGM and
the diffuse extragalactic CR flux takes over. 

\item{}
The galactic CRs which escape into the IGM have a spectral index $p\approx
2.2$ below the knees, and $p\approx 2.7$ above it. At ultra-high energies
(UHE), the spectrum of the extragalactic CRs is modified by their
interaction with the microwave background radiation (mainly, $\pi$
production, pair production and photo-dissociation of nuclei) and by the
Hubble expansion during their residence time in the IGM.

\item
The energy deposited in the IGM by the CBs and by their
associated CR jets stirs it up and generates the IGM magnetic field. 

\item{}
The CR electrons, which are accelerated/deposited directly in the IGM,
produce an extragalactic diffuse gamma-ray background radiation (GBR) via
inverse Compton scattering of the cosmic microwave background (MBR).

\end{itemize}

\noindent
\section{Collimated jets and relativistic beaming} 
\noindent
Radio, optical and X-ray observations with high spatial
resolution indicate that relativistic jets, which are fired by quasar and
microquasars, are made of a sequence of plasmoids (cannonballs) of
ordinary matter whose initial expansion (presumably with an expansion
velocity similar to the speed of sound in a relativistic gas) stops
shortly after launch (e.g., Dar \& De R\'ujula 2003 and references
therein). The turbulent magnetic fields in such plasmoids gather and
scatter the ionized ISM particles on their path. For the sake of
simplicity, we shall assume\footnote{Acceleration in the CB's rest
frame and deviation from isotropic scattering have been studied by A. De
R\'ujula, 2004, in preparation} that a fraction of the incident ISM
particles are scattered isotropically by the CB and maintain their energy
in the CB's rest frame.  Electrons which are trapped in the CB cool there
quickly by synchrotron emission.  This radiation which is emitted
isotropically in the CB's rest frame is beamed by the relativistic bulk
motion of the CBs (Lorentz factor $\gamma=1/ \sqrt{1-\beta^2}\, )$. Let
primed quantities denote their values in the plasmoid's rest frame and
unprimed quantities their corresponding values in the lab frame. Then the
angle $\theta'$ of the emitted photons in the CB's rest frame relative to
the CB's direction of motion, and the corresponding angle $\theta$ in the
lab frame, are related through: 
\begin{equation}
  \cos\theta' = {\cos\theta-\beta \over 1-\beta\, \cos\theta}\, .
\label{thetaprime}
\end{equation}
This relation is valid to a good approximation also for the emission of 
highly relativistic massive particles. When applied to 
an isotropic distribution of emitted particles 
in the CB's rest frame, it yields a distribution, 
\begin{equation}
{dn\over d\Omega}={dn\over d\Omega'}\, {dcos\theta'\over dcos\theta}
\approx {n\over 4\, \pi}\, \delta^2
\label{dist}
\end{equation}
in the lab frame where
\begin{equation}
  \delta = {1\over \gamma\, (1-\beta\, \cos\theta)}
\label{Doppler}
\end{equation}
is the Doppler factor of the CB motion viewed from a lab angle $\theta$. 
For plasmoids with highly relativistic bulk motion Lorentz factor, 
$\gamma^2 \gg 1\, ,$  and for $\theta^2 \ll 1$, 
the Doppler factor is well approximated by 
\begin{equation}
\delta \approx {2\gamma\over 1+\gamma^2\, \theta^2}\; .
\label{Dopplerapprox}
\end{equation}
Hence, the isotropic distribution of the emitted particles in the CB's rest 
frame is collimated into a narrow conical beam, ``the beaming cone'', 
around  the direction of motion of the CB in the lab frame,
\begin{equation}
{dn\over d\Omega}\approx {n\over 4\, \pi}\, \left[ {2\gamma\over 
1+\gamma^2\, \theta^2}\right]^2\, .
\label{distapprox}
\end{equation}
The beaming depends only on the CB's Lorentz factor and not on the mass of 
the scattered particles. 

\noindent
Ambient ISM particles 
which are practically at rest in the ISM, enter the CBs with an energy
$E'=\gamma\, m\, .$ After magnetic scattering and isotropization in the 
CB they are emitted with a lab energy 
\begin{equation}
E=\gamma\, E'\, (1+\beta^2\, cos\theta')\, .
\label{CRE}
\end{equation}
Their energy distribution in the lab frame is given by a simple step 
function:  \begin{equation}
{dn\over dE}= 
{dn\over dcos\theta'}\, {d\cos\theta'\over dE}\approx {n\over 2\, 
\beta^2\,\gamma^2\, m}\,\Theta(E-2\,\gamma^2\, m)\, , 
\label{Edist}
\end{equation}
where $\Theta(x)=1$ for $x<1$ and  $\Theta(x)=0$ for $x>1\, .$  
The accelerated nuclei ($m=A\, m_p$) and electrons ($m=m_e$), which are
initially ejected into a narrow cone, are later scattered and isotropized 
by the
galactic magnetic fields. They do not reach far away from their injection
cone before they radiate most of their initial energy via synchrotron
emission and inverse Compton scattering of the microwave background
photons  along their direction of motion. Their radiation, which is beamed
along their motion, can be seen by an observer outside the CBs' beaming
cone, only when their direction of motion points towards him. 
Because of the fast radiative cooling of energetic electrons, their
radiation is visible mainly when they are still within/near the beaming
cone. The gradual increase of the opening angle of the injection cone due
to jet deceleration, together with the finite lifetime of the radiating
electrons which confines them to near the jet, produce the
conical images of radio and optical jets. It is often confused
with the true geometry of the relativistic jet, which 
reveals itself only in observations at much higher frequencies 
where the observed emission requires much 
stronger magnetic fields than those present in the ISM and IGM.

\section{CR acceleration by decelerating jets}
\noindent
Let $n_{_A}$ be the density of nuclei of atomic mass $A$ along the jet
trajectory. Let us assume that the elemental abundances
$X_A=n_{_A}/n_b$ are constant along the jet trajectory, 
where $n_b$ is the total baryon density. 
Let us define a total effective mass of these particles, \,$\bar{m}=m_p\,
\Sigma A\, X_A\, .$ Let us assume that most of them are swept into the jet. 
Let us also neglect the small energy radiated away by the swept up
electrons except for the fact that it ionizes completely the ISM in front
of the jet. Then, energy conservation implies that 
$M(x)\, \gamma(x)=M_0\,\gamma_0$ where $M(x)$ and $\gamma(x)$ are the 
total rest-mass and
bulk-motion Lorentz factor of the jet along its trajectory with initial
values, $M_0=M(0)$ and $\gamma_0=\gamma(0)\, ,$ respectively. Let us
denote by $dN_{_A}(x)$ the number of cosmic ray nuclei of atomic mass $A$
which are accelerated promptly by isotropic scattering in the CB rest
frame at distance $x\, .$ Noting that their average lab energy is $
\gamma^2\, A\, m_p\, ,$ one can write the approximate deceleration law due
to CR acceleration as:  
\begin{equation} M\, d\gamma = - dN_b\, \bar{m}\, \gamma^2\,, 
\label{dec1} 
\end{equation} 
or 
\begin{equation}
dN_A\approx {X_A\, M_0\, \gamma_0 \over \bar{m}}\, {d\gamma\over
\gamma^3}\, . 
 \label{dec2} 
\end{equation} 
Note that Eq.~(\ref{dec1}) depends neither on the geometry of the jet 
nor on the density profile along the jet trajectory.
The CR energy spectrum generated by a decelerating jet can be obtained by 
replacing $n_p$ in Eq.~(\ref{Edist}) by $dn_{_A}$ and integrating over 
$\gamma$ at a fixed $E$ under condition~(\ref{CRE}): 
\begin{equation} 
{dN_A\over dE}=
          {X_A\, M_0\, \gamma_0 \over 2\, A\, m_p\, \bar{m} }\,
   \int {d\gamma\over \gamma^5}\propto {X_A\over A\, \bar{m}}\,
 \left[\left[{E\over A\, m_p}\right]^{-2}-
 \left[{E_{max}\over A\, m_p}\right]^{-2}\right]\,\Theta(E-E_{max}), 
\label{CRSbk}
\end{equation}
where, 
\begin{equation}
E_{max}=2\, A\, m_p\,\gamma_0^2 
\label{Emax}
\end{equation}
is the maximum energy gained
by nuclei at rest in the ISM, in  {\bf a single scattering}.
Thus, for energies well below $E_{max}$, the injected spectrum of 
CRs is a simple power-law, $dN_A/dE\sim E^{-2}$.   
The effective power on the rhs of Eq.~(\ref{dec1}) may be increased slightly
by the scattering of ISM particles by the ambient magnetic field which is
swept up by the CRs that were scattered by the CB (to a mean Lorentz
factor $\gamma^2\, ).$ Then the  mean Lorentz factor of these secondary
accelerated particles is $\gamma^4\, .$ Such a multiple acceleration has 
been assumed to take place within relativistic shocks in collisionless
shock acceleration. A modified acceleration law,
\begin{equation}
M\, d\gamma = - dN_b\,\bar{m}\, \gamma^{2.4}\, ,
\label{decp}
\end{equation}
is needed in order to generate a slightly steeper power-law spectrum,
$dN_A/dE\sim E^{-2.2}\, ,$ which was deduced from the spectrum of the
diffuse gamma ray emission by Galactic cosmic ray electrons (e.g. Dar and
De R\'ujula 2001a) and from the diffuse radio emission from cosmic ray
electrons in our galaxy, in external galaxies and in clusters of galaxies. 
It has been claimed that such a power-law index 
arises in numerical calculations
of CR acceleration in collisionless shock acceleration (Bednarz \&
Ostrowski 1998; Kirk et al.~2000).  In the following, we shall assume that
the injection spectrum of CRs by relativistic jets has a spectral index
$p\approx -2.2\, .$

\section{Spectral steepening by magnetic trapping} 
\noindent
The cosmic rays which are accelerated by the highly relativistic jets from
SN explosions are initially beamed into narrow cones along the jets'
trajectories. Their free escape into the intergalactic space is delayed by
diffusion in the galactic magnetic fields which isotropize their direction
of motion. The accumulation of CRs during their trapping/residence time in
the galaxy, which decreases with increasing energy, steepens their energy
spectrum.  During their galactic residence, CRs also lose energy via
Coulomb and inelastic collisions and gain energy through Fermi
acceleration by galactic winds and jets.  We shall first neglect
reacceleration of CRs in the ISM and their inelastic interactions there,
and correct only for accumulation during residence time. 

\noindent
The Larmor radius of CRs with an electric charge $Z$ 
in a magnetic field $B$ is given by
\begin{equation}  
R_L={\beta\, E\over e\, Z\, B}\approx \left[ {E\over PeV}\right]\, 
    \left [{1\over Z\, B_{\mu G}}\right]\, pc\,,      
\label{Larmor}
\end{equation}
where $B_{\mu G}$ is the magnetic field strength in $\mu Gauss$. The
energy-dependent diffusion of CRs due to pitch-angle scattering with a
Kolmogorov and Kraichnan spectrum of MHD turbulence results in a CR
residence time in the host galaxy which behaves (e.g. Wick, Dermer
\& Atoyan 2003) like $ (E/Z)^{-0.5}\,.$ Thus, the accumulation of CRs before
their escape steepens the injection spectral index, $p=-2.2\,,$ by 0.5 to
$p=-2.7$, yielding,
\begin{equation}
{dN_A\over dE} \propto {X_A(E/A)A^{1.2}Z^{0.5} \over \bar{m}} 
                      \left[{E\over m_p}\right]^{-0.5} 
                      \left[\left[{E\over  m_p}\right]^{-2.2}-
                      \left[{E_{max}\over m_p}\right]^{-2.2}\right]
                      \Theta(E-E_{max}). 
\label{CRdnde} 
\end{equation} 
The decreasing metalicity and the  deceleration of the jet 
along the jet's trajectory  result in  an effective energy dependent  
$X_A(E/A)$ which is discussed in section 7. 
Eq.~(\ref{CRdnde}) ignores acceleration inside the CBs and 
reacceleration of CRs in the ISM by the CBs.
Such effects will be included phenomenologically in section 9.

\section{CRs from SN jets}
\noindent
There is mounting observational evidence that long duration gamma ray
bursts (GRBs) are produced by ultra-relativistic jets of ordinary matter
which are ejected in core collapse supernova (SN) explosions (see e.g.
Dar 2004; Dado these proceedings) as long advocated by the remarkably
successful CB model of GRBs (e.g., Dar \& De R\'ujula 2000, 2003; Dado et
al.~2002, 2004 and references therein). In the CB model, 
the long-duration GRBs are produced in ordinary core-collapse SN 
explosions. Following the
collapse of the stellar core into a neutron star or a black hole, and
given the characteristically large specific angular momentum of stars, it 
is hypothesized that an
accretion disk or torus is produced around the newly
formed compact object, either by stellar material originally close to the
surface of the imploding core and left behind by the explosion-generating
outgoing shock, or by more distant stellar matter falling back after its
passage (De R\'ujula 1987). A CB is emitted, as observed in microquasars,
when part of the accretion disk falls abruptly onto the compact object
(e.g.~Mirabel \& Rodrigez 1999; Rodriguez \& Mirabel 1999 and references
therein). The high-energy photons of a single pulse in a GRB 
are produced as a CB coasts
through the ``ambient light'' permeating the surroundings of the
parent SN. The electrons enclosed in the CB Compton
up-scatter photons to energies which, close to the CBs direction
of motion, correspond to the $\gamma$-rays of a GRB and
less close to it to the X-rays of an XRF.
Each pulse of a GRB  corresponds to one
CB in the jet. The timing sequence of emission of the successive individual 
pulses (or CBs) reflects the chaotic accretion process and its
properties are not predictable, but those of the single pulses are (Dar \&
De R\'ujula 2003 and references therein). The initial Lorentz factors
of the CBs were deduced from
cannonball model analysis of GRBs and their afterglows (Dar \& De 
R\'ujula~2003). Their 
distribution,
and the corresponding distribution of $E_{max}\,,$ can be well approximated 
by a log-normal distribution:
\begin{equation}
P(E_{max})\approx  {1\over E_{max}\,\sigma\, \sqrt{2\, \pi}}\, 
   exp\left[-{(lnE_{max}-ln\bar{E}_{max})^2\over 2\, \sigma^2}\right]\, . 
\label{PEmax}
\end{equation}
The energy-spectra of the individual CR elements
is obtained by integrating Eq.~(\ref{CRdnde}) over $E_{max}$ with the 
log-normal distribution~(\ref{PEmax}):
\begin{equation}
{dN_A\over dE}\rightarrow \int P(E_{max})\, {dN_A\over dE}\, dE_{max}\, . 
\label{CREspectrum}
\end{equation}
So far we have ignored acceleration inside the CBs and
reacceleration of CRs in the ISM by the CBs.
Such effects will be included in the following sections.

\section{The CR knees}
\noindent
The log-normal distribution of the Lorentz factors of 
CBs, as inferred from GRBs (Dar \& De
R\'ujula~2003), produce a relatively narrow distribution of the maximum 
energy of CRs produced by elastic magnetic scattering of ambient ISM 
particles in the CB. 
Thus, the energy spectra of the individual CR nuclei, integrated over all 
CBs, retain a sharp knee at 
\begin{equation}
         E_{knee} \approx \bar{E}_{max}= 2\, A\, m_p\, \bar {\gamma_0^2} 
                    \approx 3\,A\, PeV\, . 
\label{Eknee} 
\end {equation}
This is shown in 
Fig.~(\ref{spectrumA}) which compares the predicted energy spectra of H, He,
and Fe group nuclei around their respective knees
as calculated\footnote{For the 
sake of simplicity,
Fermi acceleration inside the CBs was  
roughly approximated by an effective $\sigma$ which was fixed 
by a best best fit 
to the proton data.} 
from Eqs.~(\ref{CRdnde}), (\ref{CREspectrum}), (\ref{Eknee}),
and those extracted from the
Kascade observations (Kampert et al.~2004; Hoerandel et al.~2004). 
Despite the large experimental uncertainties, the theory seems to 
reproduce correctly the elemental knees, their A-dependence  
and their energy spectrum around these knees.
Note that
the predicted knee for different elements is proportional to $A$ rather than
to $Z$
as in conventional models based on shock acceleration, where
the maximum energy gain is limited by the requirement that the size
of the accelerator be larger than the Larmor radius of the accelerated
CR particles. This difference is 
large only for protons, deuterium and very heavy nuclei. 
It appears to be supported by the data, but more accurate
values
for the knee energy of different nuclei are needed 
from the CR experiments in order to draw a reliable conclusion.

\section{Elemental abundances and spectral indices below the knees}
\noindent
The enhancement in the abundance of CR nuclei 
as function of nuclear mass, charge and CR energy below the elemental knee, 
which is predicted by the CB model, can be read from 
Eq.~(\ref{CRdnde}), 
\begin{equation}
X_A[CR] \sim  X_A(E/A)\, A^{1.2}\ Z^{0.5}\, ,
\label{CRabund}
\end{equation}
where $X_A(E/A)$ is roughly the element's abundance encountered 
on the average by the CBs 
along their trajectory in the ISM around a distance  where the CB Lorentz 
factor is $\gamma\approx E/A\,  m_p\, .$ The $A$-dependence follows 
from the fact that all CR nuclei, which
are accelerated by CBs, have the same universal Lorentz factor 
distribution. The $Z$-dependence follows from the 
dependence of Galactic magnetic-trapping time of CRs on their Larmor radius.

\noindent
More than 90\% of SN explosions take place in star formation-regions which
are enclosed in superbubbles (SB) formed by the ejecta from former SN
explosions and massive stars.  Therefore, near the GRB site, 
the elemental abundances $X_A$ are
typically that of young SNRs (which are made of SN ejecta +
progenitor ejecta prior to the SN). Further down  they become typical SB
abundances. Both $X_A[SNR]$ and $X_A[SB]$ are poorly known.  
Outside the SB, they become normal ISM abundances and when the
jet enters the halo, $X_A$ become typical halo abundances. 
The decrease in metalicity 
along the CB trajectories, approximately by a factor of a few,   
induces a noticeable change 
in the spectral index of the different CR elements. 
It makes the proton spectrum slightly steeper and 
decreases slightly the steepness of the energy spectrum of the metals. 
The change in the spectral index, 
is given roughly by 
\begin{equation}
p_{_{A}}\approx 2.7+{\Delta [log{X_A[E/A]}]\over
                   \Delta [log(E/A)]}\, . 
\label{steep} 
\end{equation}
The change $\Delta$ in elemental abundances along the CB
trajectory from $X_{A}[ISM]\approx X_A[\odot]$ (given e.g. by Grevesse \& 
Sauval 1998) to $X_{A}\sim X_A[SNR] $
corresponds to an energy-increase from $E\sim m_{_A}$ to $E\sim
E_{max}[A]\,.$ 
Thus, for protons, $p_{_1}\approx 2.75$ below the proton knee, 
while for iron nuclei below the iron knee,
$p_{_{56}}\approx 2.62\,.$ In Fig.~\ref{pA} we compare the
predicted spectral index below the knee for different elements and their
observed values as reported by Wiebel-Sooth et al.~(1998) from their best
fits to the world CR data below the knee. The CB model
prediction for all the abundant elements above $He$ is described 
approximately by the interpolation, $p_{_A} \sim 2.70-0.02\, lnA\, .$  

\noindent

During their residence time in galaxies, spallation of CRs in collisions
with ISM nuclei increases significantly the abundances of long-lived rare
elements, but changes only slightly the abundances of the most abundant
elements. A detailed discussion of the effects of spallation on CR
elemental abundances is beyond the scope of this paper.  Thus, Table I
presents the observed CR elemental abundances near TeV energy and
$K_A[SB]=X_A[SB]/X_A[\odot]$ the enhancement in the abundance of the the
most abundant elements in SBs relative to their solar abundances, which is
needed in the CB model to explain the observed CR abundances. Indeed,
these enhancements are consistent with an expected SB metalicity $\sim 5$
times solar (Lingenfelter et al.~2001) and with those extracted from X-ray
spectra of SBs measured by XMM newton (with large uncertainties).  Table I
also lists the values of $X_A[ISM]\approx X_A[\odot]\,$ which we used in
the CB model calculations. Note the large enhancement in the abundance of
helium and the huge enhancements in the abundances of the metals in CRs
compared to their ISM and solar values. Ignoring spallation, the CB model
reproduces very well these very large enhancements. 

\noindent 
Above the proton knee, CRs become progressively poor in protons. Above the
$He$ knee, they become poor also in $He\, ,$ then also in $CNO$, etc.,
until near the iron knee (the second knee) where they consist mainly of
iron and heavier elements.  Spallation affects only slightly the value of
$<lnA>$. This is shown in Fig.~\ref{lnA} where the CB model 
prediction for
$<lnA>$  as a function of CR energy is compared with its value 
as inferred from various observations (see, e.g. Hoerandel).  
The effects of spallation were neglected in the CB model predictions.
Despite the large spread
in the world data, the CB model clearly reproduces the observed trends. 

\noindent
The energy spectrum of individual elements falls rapidly at their knees.
The all-particle CR spectrum beyond the proton knee loses progressively
the contribution from heavier and heavier elements, until it becomes
almost pure iron. This produces 
a steepening of the spectrum between
the proton knee and the iron knee and results in a composition which
gradually approaches a pure iron+heavier metals composition
as shown in Fig.~\ref{lnA}.
Because the
abundances of elements heavier than iron are rather small compared to
iron, {\bf the all-particle spectrum steepens again at the iron knee 
forming the
second knee} in the all-particle spectrum. Beyond the second knee,
reacceleration of CRs in SBs become the main sorce of CRs.

\noindent 
\section{The maximal energy of reaccelerated CRs} 
\noindent 
A cosmic ray in the ISM with an energy  $E_L$ which
collides with a CB at angle  $\theta_{_L}$ relative to the 
direction of motion of the CB, has an
energy $E'=\gamma\, E_L\, (1-\beta\, \beta_{_L}\, cos\theta_{_L})$   
in the CB rest frame. If it is 
scattered elastically and emitted at angle $\theta$' 
in the CB rest frame, its energy in the ISM rest frame becomes
\begin{equation}
E=\gamma^2\,E_L\, (1+\beta'\,\beta\, cos\theta')\, (1-\beta\,\beta_{_L}\,  
           cos\theta_{_L})\,. 
\label{CREDS}
\end{equation}
However, CBs can accelerate CRs as long as 
their Larmor radius in the 
CB rest frame is smaller than the CB's radius, $R_{cb}\,,$ i.e., $E'$ must 
satisfy, $ E'\leq e\, Z\, B\, R_{cb}\, .$ Hence, the maximal energy 
of CR nuclei accelerated directly by/in CBs is 
\begin{equation}
E_{max}\approx 4\, \gamma\,e\, Z\, B\, R_{cb}\,.
\label{maxds}
\end{equation}
This relation differs from the usual `Hillas relation' by the factor $
4\,\gamma$ which is due to the relativistic motion of the CR accelerator 
-- the CB.  For a magnetic field whose pressure is equal to that 
of the scattered ISM particles (Dado et al.~2002),
$B\approx \gamma\,\sqrt{2\, \pi\, n_p\, \bar{m}}\,,$ 
where  $n_p$ is the superbubble density, the maximal energy is,
\begin{equation}
E_{max}[A]\approx 4\times 10^{20}\, Z\, \left[{\gamma_0^2\over 10^6}\right]\,
       \left[{n_p\over 10^{-3}\, cm^{-3}}\right]\,
        \left[{R_{cb}\over 10^{14}\,cm} \right]\, eV\, .
\label{maxdsA}
\end{equation}
Hence, SN jets which produce GRBs can accelerate CR nuclei to energies 
much higher than the highest energy of a cosmic ray  which has ever been 
measured, $\sim 3.2\times 10^{20}\, eV$ (Bird et al. 1995).  

\noindent
\section{Reacceleration of CRs by CBs} 
\noindent
CRs which are injected into the ISM and IGM by  SN jets 
can be reaccelerated later by 
magnetic scattering from highly relativistic jets (CBs) from
other SN explosions. This reacceleration could have been
taking place since the beginning of star formation.
Detailed treatment of reacceleration is beyond the scope of this paper.
A rough estimate of
the energy spectrum of CRs resulting from reacceleration of CRs 
in SBs by magnetic scattering from CBs can be obtained by following the 
calculation of the
spectrum of CRs produced by CBs through magnetic scattering of ISM 
particles at
rest.  One has to replace  $m$ in Eq.~(\ref{CRE}) by $\gamma\, 
E_L\, (1-\beta\, \beta_{_L}\, cos\theta_{_L})$ and integrate over the CR
spectrum as given by Eqs.~(\ref{CRdnde}),~(\ref{CREspectrum}) and over 
$cos\theta_L$, subject to Eq.~(\ref{CREDS}). An approximate integration 
yields a population of reaccelerated CRs above the CR knees with an 
energy spectrum,
\begin{equation} 
   {dN_A\over dE} \propto \left[{E\over 
m_p}\right]^{-p_{_A}}\, , 
\label{CRdnded}
\end{equation}
which extends up to $E\approx E_{max}[A]$ given by Eq.~(\ref{maxdsA}). 
Because of
the steep decline with energy, $\sim E^{-2.7}\,,$ of the CR spectrum below 
the CR
knee, most of the CRs nuclei with energy $E>E_{knee}$ were reaccelerated
in SBs from an initial CR energy $\sim E/\gamma_0$ to their final energy 
$E\, .$ Because the energy $E$ of reaccelerated CRs, 
depends only on the 
energy in the CB rest frame, and not  
on their mass or charge (as long as 
$E<E_{max}[A]\, ,$)
the composition of reaccelerated CRs is approximately the CR
composition at the all-particle CR knee.

\noindent 
\section{The CR ankle and beyond}
\noindent 
\noindent
In a steady state, the bulk of the extaragalactic CRs are injected into the 
IGM with their galactic injection spectrum i.e.,  a 
power-law 
spectrum with  a spectral index $p=p_{_A}-0.5$ below the elemental knee  
and $p=p_{_A}$ above it, all the way to $E_{max}\, .$  
During the Hubble 
time, this spectrum is
modified by the cosmic expansion and by pair production, photoproduction
and photodissociation on the MBR photons (see e.g., Stecker \& Salamon 
1999).

\noindent
In the CB model the CR ankle is the energy where the residence time of
cosmic rays in the Galaxy approaches the free escape time. This happens
when the Larmor radius of the CRs approaches the coherence length of the
turbulent Galactic magnetic fields (Dar \& Plaga 1999).
Hence, the elemental ankles satisfy $E_{ank}[A]=Z\, E_{ank}[p]\,.$
The exact value of $E_{ank}[p]$ cannon be calculated from first principles
and is traeted as an adjustable parameter. Beyond the ankle,
the CR energy spectrum is dominated by the flux of the CRs from the IGM
which have been injected there by SN explosions in galaxies
in the local super cluster since the
beginning of star formation, and were isotropized  by the IGM magnetic
fields.  The contribution of CR-reacceleration to the observed
elemental CR flux, well above the knee, can be written as,
\begin{equation}
   {dN_A\over dE} \propto
    {X_A[SB]\, A^{1.2}\, Z^{0.5}\over \bar{m}}\, 
   \left[{E\over m_p}\right]^{-p_{_A}}\, 
   \left( \left[E\over E_{ank}[A]\right]^{-0.5}+1\right)\, exp(-E/E_0[A])\, ,
\label{CRdndereac}
\end{equation}
where $E_0[A]\approx E_{GZK}[A]\sim  5\times 10^{19}\,A\, eV$
is the effective CR cutoff due to photo-production of $\pi$
(the GZK cutoff) or photo-dissociation of nuclei (Puget, 
Stecker \& 
Bredekamp, 1976; Stecker \& Salamom 1999) on the MBR photons. Below 
$E_{max}[A]$,  reacceleration  is
blind to $A$ and $Z$. Consequently, the proportionality constant in
Eq.~(\ref{CRdndereac}) is independent of $A$ and $Z\, .$
At the ankle the CRs residence time in the Galaxy is approximately their
free escape time from the Galaxy. This is described by the factor
$ [(E/E_{ank})^{-0.5}+1]$ on the rhs of Eq.~(\ref{CRdndereac}).
The accumulation time of CRs in the IGM
is roughly the age of the Universe, $t_{_H}\sim 14\, Gy\,.$
The average IGM volume  in the local Universe per
Milky Way-like
galaxy is $\bar{V}\sim 100\, Mpc^3\, ,$ while the volume of the
Galactic  CR halo is roughly (Dar \& De R\'ujula 2001b)
 $\bar{V}\sim 10^5\, kpc^3\,.$
Thus, ignoring redshift and stellar evolution (which tend to compensate
each other), the extragalactic
(EG) flux near the CR ankle must satisfy roughly,
\begin{equation}
   {dN_{EG}\over dE} \sim  {t_{_H}\over t_{esc}}\,
                          {V_{G}\over \bar{V}}\,  {dN_{G}\over dE}
                     \sim {dN_{G}\over dE}\, . \label{EGCR} \end{equation}
This relation is nominally satisfied if the mean escape time of CRs from
the Galaxy at the ankle is, $t_{esc}\approx 13,500\, y\, .$
\noindent
The spectral index of UHECRs  above the ankle can be predicted also
from a general consideration: Below the ankle, magnetic trapping 
increases the CR index by 0.5. The measured spectral index below the 
ankle is $p\approx 3.2\, .$ Thus, free escape of CRs changes it to 
$p\approx 3.2-0.5=2.7$ at the ankle.

\noindent
In conclusion, in the simple version of the CB model, reacceleration 
leads to four major predictions:
\begin{itemize}
\item{}
The elemental spectral index of UHECRs above the ankle is identical to that
below the knee, $p_{_A}\, $ and consequently the all-particle spectral index
above the all-particle ankle is approximately  2.7 .
\item{}
The elemental composition of UHECRs above the ankle and below
the effective 
energy thresholds for photoproduction/photodissociation 
in collisions with the MBR photons 
is similar to that of
CRs below the CR knee.
\item{}
The CR spectrum steepens at the GZK cutoff.
\item{}
The CR spectrum above the ankle remains highly isotropic
up to the GZK cutoff.
\end{itemize}

\noindent
The all-particle CR spectrum between $10^2$ GeV and $10^{12}$ GeV as
predicted by the CB model is compared in Fig. 1
with the world data as compiled by Ulrich (Kampert et al.~2004). Despite
the large uncertainties in the experimental data, and in the model input
parameters (e.g. SB abundances, photo-nuclear cross sections), the agreement 
between theory and experiments appears to be quite good. Note in particular 
that the predictions above $10^{11}\,  GeV$ are sensitive to the assumed
photo-nuclear cross sections, cosmological model and past SN rates
in the local supercluster.

\noindent 
\section{The Galactic CR Luminosity}
In a steady state, the escape rate of Galactic CRs into the IGM 
is equal to their production rate. In the CB model, most of the kinetic 
energy of SN jets is converted to CR energy in the Galaxy. 
In the CB model, this energy  was estimated {Dar \& De R\'ujula 2000, 2003)
to be $E_k\sim 2\times 10^{51} erg$ per SN. Thus, a galactic SN rate of 
$R_{SN}\sim 1/50\, y^{-1}$ generates a Galactic CR luminosity of 
\begin{equation}
   L_{cr}[MW]=R_{SN}\, E_k\sim 1.1\times 10^{42}\, erg\, s^{-1}. 
\label{EGCR}
\end{equation}
\noindent
Traditional estimates, which are based on an assumed $\sim 10\%$ conversion 
of the kinetic energy of the non-relativistic ejecta in SN explosions  
to CR energy through collisionless shock acceleration, give $L_{cr}< 
10^{41}\, erg\, s^{-1}\,.$

\noindent 
\section{IGM magnetic field}
\noindent 
In the CB model, practically all the kinetic energy of the SN jets (CBs), 
$E_k\sim 2\times 10^{51}\, ergs$ per SN  explosion  ends up in the 
IGM (Dar \& Plaga 1999; Dar \& De R\'ujula 2000,2003).
The energy and momentum deposited in the IGM  may stir it up and  
generate 
the IGM magnetic fields. If their energy is equipartitioned with the 
IGM plasma (primordial matter and galactic winds) and the IGM magnetic 
fields then, neglecting cosmic-evolution corrections, the energy density of 
these fields is roughly given by,
\begin{equation}
B\sim \left[{8\, \pi\, R_{SN}\, E_k\over 3\, H_0}\right]^{1/2} \approx 20\, 
          nGauss\, ,  
\label{IGMB}   
\end{equation} 
where $R_{SN}\approx 6\times 10^{-5}\, Mpc^{-3}\, y^{-1}$ is the total rate 
of SN explosions per unit volume in the 
local universe and $1/H_0\approx 14\, Gy$ is the approximate age of the 
Universe. If only SNIb/Ic produce relativistic jets/GRBs  
then $B\sim 10\,nGauss\, .$
These rough estimates are consistent with estimates based on 
measurements of Faraday rotation of polarized light from distant quasars. 

\section{The extragalactic GBR produced by CRe}   
\noindent
Inverse Compton scattering of CR electrons (CRe) in external 
galaxies and in the IGM produces  
an isotropic GBR with a canonical spectrum 
$dn_\gamma/dE\sim E^{-2.1}$ whose magnitude has been estimated 
in Dar \& De R\'ujula 2001a to be less than $40\%$ of the diffuse 
gamma ray background radiation at high Galactic latitudes. This 
component should be detected by GLAST. Such an extragalactic 
component will reduce   
the  CR luminosity of the Milky Way Galaxy,
   $L_{cr}[MW]= 6.8\times 10^{42}\, erg\, s^{-1}\,,$
as estimated by Dar \& De R\'ujula (2001), by a factor of $[0.6]^{3}$ 
to the value  $L_{cr}[MW]=1.5\times 10^{42}\, erg\, s^{-1}\,, $ 
similar to that estimated in Eq.~(\ref{EGCR}). 

\section{conclusions} 
\noindent 
The CB model of GRBs was used to show that the bipolar jets from SN
explosions are the main origin of cosmic rays at all energies.  The CB
model correctly predicts, within the experimental uncertainties, the
observed all-particle CR flux, its energy spectrum and its elemental
composition as function of energy.  At ultra-high energies above the
CR ankle, for which the Galactic magnetic fields can no longer delay the
free escape of CRs from the Galaxy, the UHECRs which were injected there 
over the Hubble time by SN jets from all the galaxies in the local 
supercluster
and were isotropized there by the IGM magnetic fields, dominate the
Galactic CR spectrum.  A GZK cutoff due to the interaction of UHECRs with
the microwave background radiation is expected. The CR nuclei which
diffuse out of galaxies, or are directly deposited in the IGM by the
relativistic SN jets, may be the origin of the IGM magnetic fields. 
Inverse Compton scattering of MBR photons by the CR electrons in the IGM
produce a diffuse extragalactic gamma-ray background radiation. It seems
that the 92 years old puzzle of the origin of Galactic CRs may have been
solved, but more precise CR data on the energy spectrum of individual
nuclei, from the elemental knees  all the way well above the GZK cutoff,
and astrophysical data on the elemental composition of SNRs and SBs, as
well as more precise CB model calculations which include CR interactions
with the ambient matter and radiation in the ISM and the IGM are needed to
secure this conclusion.

{\bf Acknowledgements:} The author would like to thank Shlomo Dado and
Alvaro De R\'ujula for a long term collaboration in the development of the
CB model for GRBs and CRs, Jacqaues Goldberg for technical help and
the organizers of La Thuile 2004 meeting and the Vulcano 2004 Workshop
for their warm hospitality, excellent organization and very interesting
scientific programs.  The support of the Asher Space Research Institute at
the Technion is gratefully acknowledged. This research was not supported
by the Israel Science Foundation for basic research.

{\footnotesize
\begin{table}[t!]
\begin{center}
\caption{Elemental abundances in the ISM and in CRs at 1 TeV 
          and their predicted  enhancement, $K_A[SB]\,,$ in 
          supernova remnants and  galactic superbubbles.} 
\bigskip
\begin{tabular}{lllll}
\hline
Element  & $X_A[\odot]$ & $X_A[CR]$ & $K_A[SB]$ & \\
\hline
H & 0.915 $\pm$0.005 & 0.433$\pm$ 0.015    &0.47  &   \\
He & 0.083$\pm$0.005 & 0.270$\pm$ 0.007    &0.43  &   \\ 
Li & $1.15\pm 0.11\,10^{-11}$ &  $7.82 \pm 0.19\, 10^{-3}$ &  & \\ 
Be & $2.26\pm\,0.23\, 10^{-11}$  & $1.78 \pm 0.18\, 10^{-3}$ & & \\ 
B  & $4.58\pm\,0.46\, 10^{-10}$ &   $3.36 \pm 0.30\, 10^{-3}$& & \\ 
C & $3.01 \pm 0.68\, 10^{-4}$   & $3.98 \pm 0.04\, 10^{-2}$ & 2.73& \\ 
N & $7.61 \pm 0.15\, 10^{-5}$   &   $8.83 \pm 0.30\, 10^{-3}$ &1.85& \\ 
O & $6.18 \pm 1.09\, 10^{-4}$   & $ 5.15 \pm 0.15\, 10^{-2}$ &1.05 &  \\ 
F & $2.89 \pm \,0.29\, 10^{-8}$  & $1.23 \pm 0.18\, 10^{-3}$ &&   \\ 
Ne & $1.098 \pm 0.18\, 10^{-4}$ & $1.73 \pm 0.37\, 10^{-2}$ &1.35&  \\ 
Na & $1.96\pm 0.16\, 10^{-6}$  & $ 2.83 \pm 0.12\, 10^{-3}$ &   &  \\ 
Mg & $3.48 \pm 0.50\, 10^{-5}$  &  $3.01 \pm 0.98\, 10^{-2}$ & 5.5&  \\ 
Al & $2.70 \pm 0.57\, 10^{-6}$ & $4.25 \pm 0.57\, 10^{-3}$   &  & \\ 
Si & $3.25 \pm 0.61\, 10^{-5}$ & $2.99\pm 0.56\, 10^{-2}$    &4.5  & \\ 
P & $2.58 \pm 0.83\, 10^{-7}$  & $1.02 \pm 0.75\, 10^{-3}$ &   & \\ 
S & $1.33 \pm 0.26\, 10^{-5}$  & $8.64 \pm 0.90\, 10^{-3}$ & 2.53 &    \\ 
Cl & $1.96 \pm 0.24\, 10^{-7}$ & $1.10 \pm 0.07\, 10^{-3}$ &    &  \\ 
Ar & $1.92 \pm 0.10\, 10^{-6}$  & $3.14 \pm 0.02\, 10^{-3}$ & 4.60 & \\ 
K  & $ 1.21\pm 0.36\, 10^{-7}$  & $2.02 \pm 0.06\, 10^{-3}$ &   & \\ 
Ca & $2.09 \pm 0.12\, 10^{-6}$  & $5.44 \pm 0.45\, 10^{-3}$ & 6.96 &  \\ 
Sc & $1.35 \pm 0.14 \, 10^{-9}$  & $1.14 \pm 0.07\, 10^{-3}$&  &  \\ 
Ti & $9.6 \pm 0.36\, 10^{-8}$  & $ 4.25\pm 0.53\, 10^{-3}$ &  &  \\ 
V & $9.15 \pm 0.92\, 10^{-9}$ &  $2.36 \pm 0.11\, 10^{-3}$  &  & \\
Cr & $4.28 \pm 0.36\, 10^{-7}$  &  $ 5.11\pm 0.45\, 10^{-3}$ &   &  \\ 
Mn & $2.25 \pm 0.20\, 10^{-7}$ &  $5.07 \pm 0.53\, 10^{-3}$ &   &  \\ 
Fe & $2.89 \pm 0.38\, 10^{-5}$  &  $6.69 \pm 0.67\, 10^{-2}$ & 3.62 &   \\ 
Co & $7.61 \pm 0.16\, 10^{-8}$  &  $2.82 \pm 0.26\, 10^{-4}$ &  &  \\ 
Ni & $1.62 \pm 0.19\, 10^{-6}$  & $3.74 \pm 0.16\, 10^{-3}$ & 3.33 &  \\ 
 
\hline
\end{tabular}
\end{center}
\end{table}
}

\begin{figure}[t!]
\vspace{10truecm}
\includegraphics{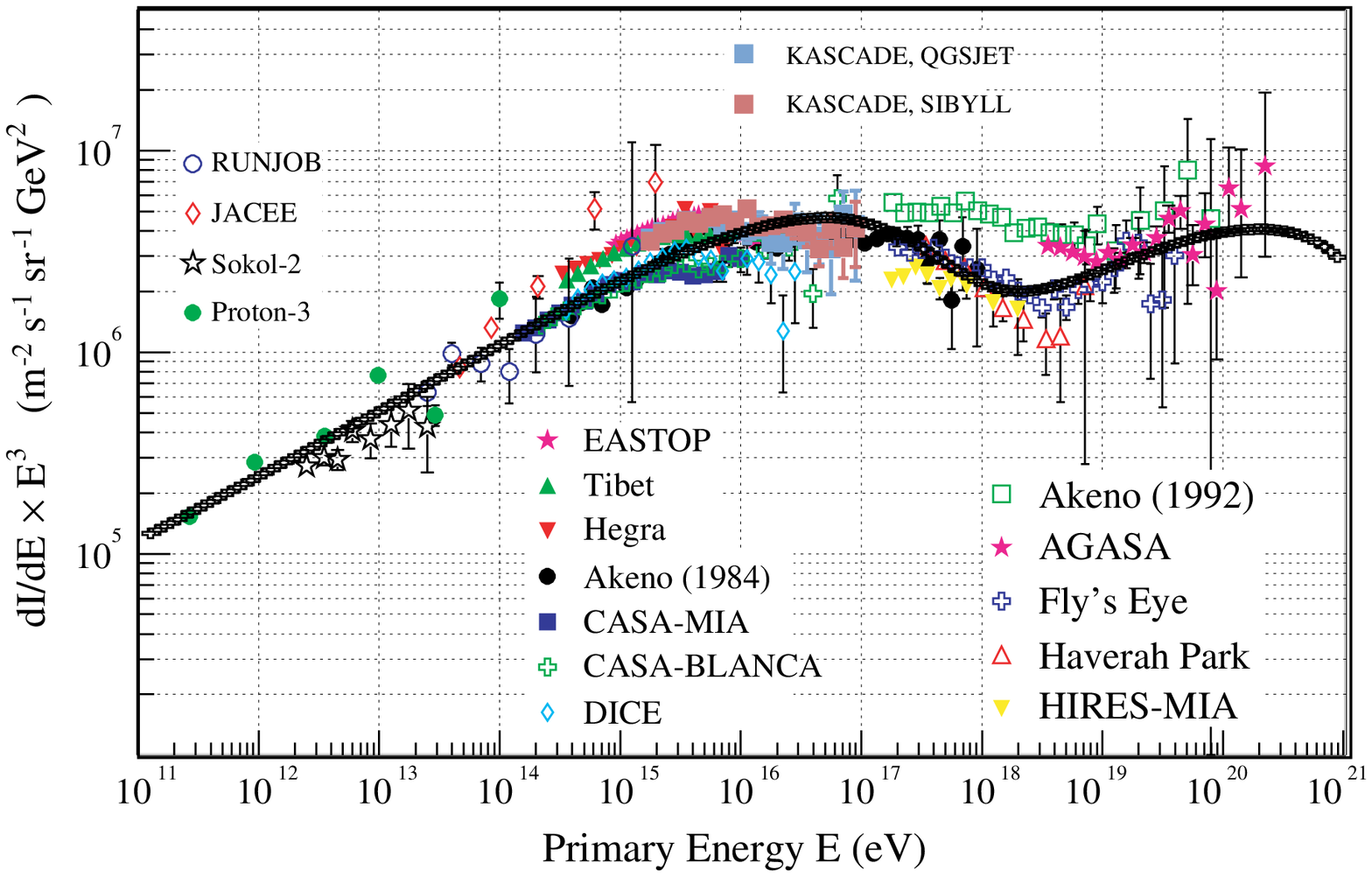}
\includegraphics{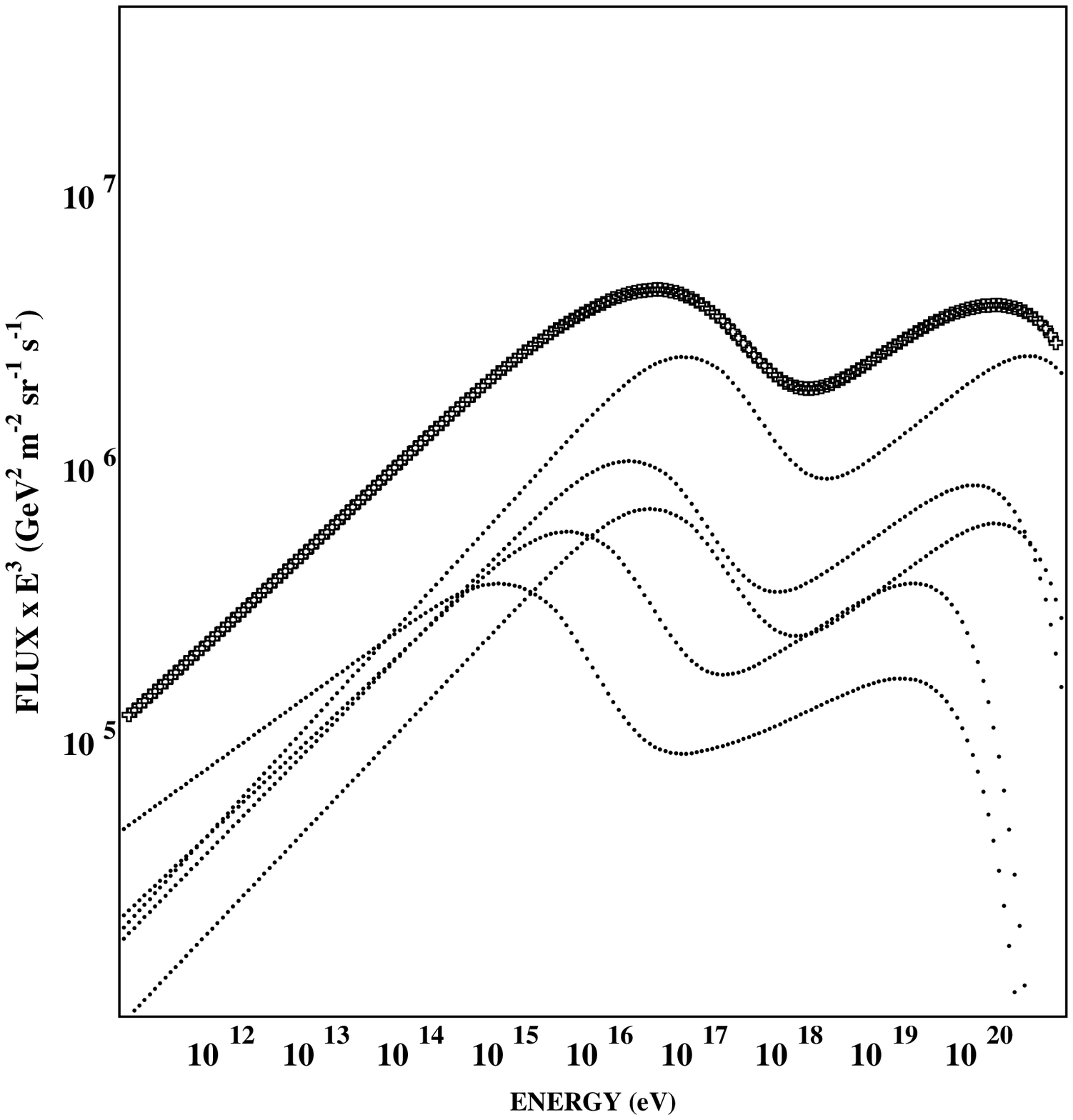}
\caption[h]{Top: Comparison between the CB model prediction for  the 
all-particle CR spectrum (crosses) and the world 
data as compiled by Ulrich (Kampert et al. 2004). 
Bottom: The break down of the CB model prediction for the all-particle CR 
spectrum (crosses) 
into the contributions from the most abundant nuclei (lhs, top to bottom): 
H, He, CNO, Ne-Ca and Fe group  nuclei.     
The theoretical 
predictions and the  observations have been multiplied in both figures by 
$E^3$ 
to emphasize significant deviations from a single power-law decline 
over thirty orders of magnitude. } 
\label{specall}
\end{figure}

\begin{figure}[t!]
\vspace{10truecm}
\includegraphics{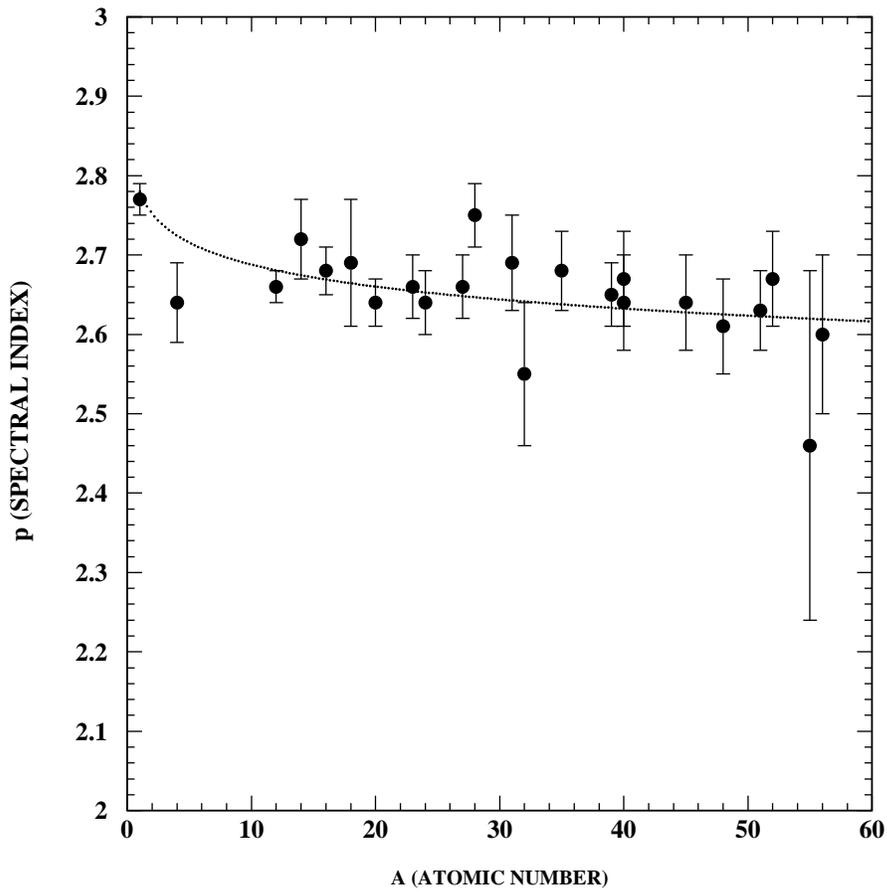}
\caption{Comparison between the CB model prediction for the 
A-dependence of the spectral index of CRs  for energies below 
the elemental CR knees, Eq.~(\ref{CRabund}), and  their values obtained by 
Wiebel-Sooth et al. (1997) from  best fits to the world CR data.}
\label{pA}
\end{figure}

\begin{figure}[t!]
\vspace{10truecm}
\includegraphics{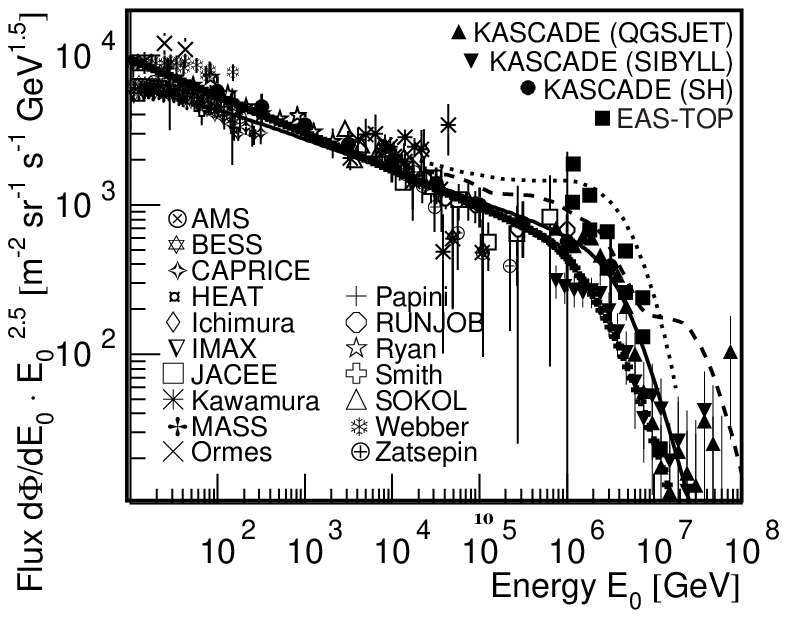}
\includegraphics{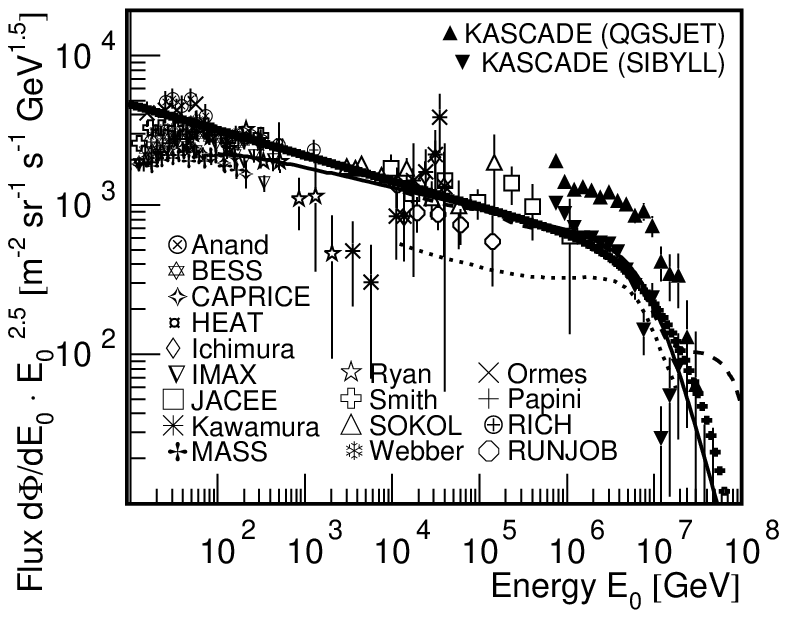}
\includegraphics{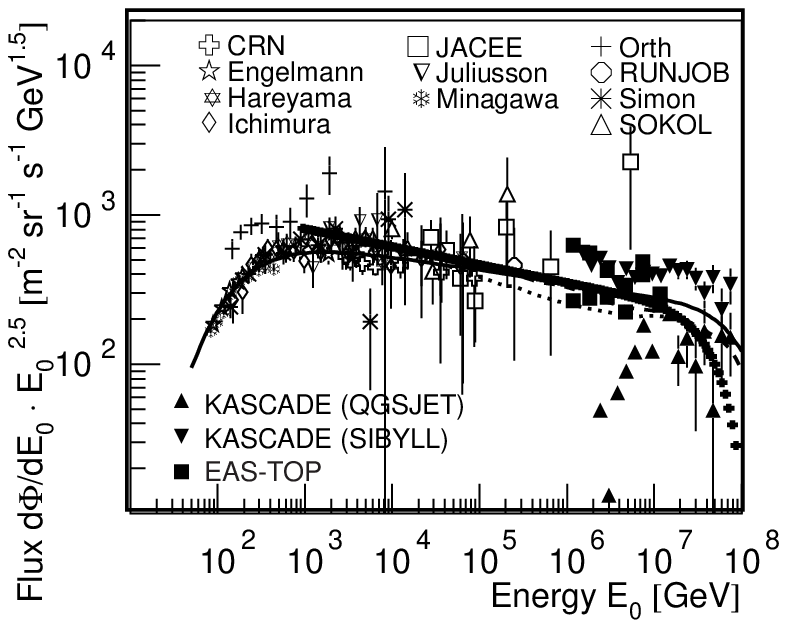}
\includegraphics{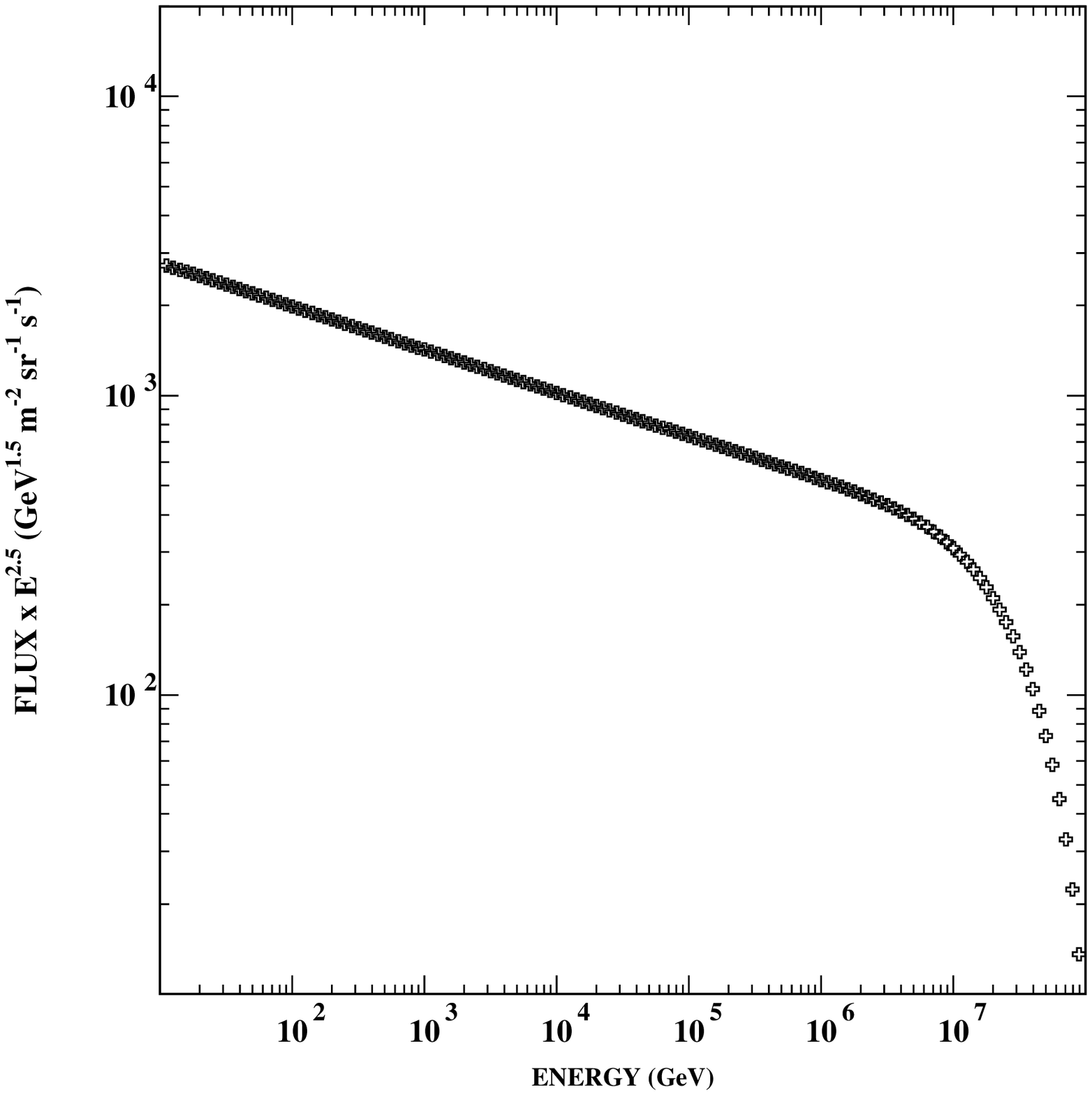}
\vspace{1truecm}
\caption{Comparison between the single-acceleration CB model predictions 
for the
energy spectrum of various cosmic ray  nuclei (thick line of crosses)
and their energy spectrum as
derived  
from various experiment and theoretical models 
and compiled by Hoerandel~(2004).
Top left -- protons, top right -- helium  nuclei, 
bottom left -- iron group nuclei and bottom right -- CNO group nuclei.}
\label{spectrumA}
\end{figure}

\begin{figure}[t!]
\vspace{10truecm}
\includegraphics{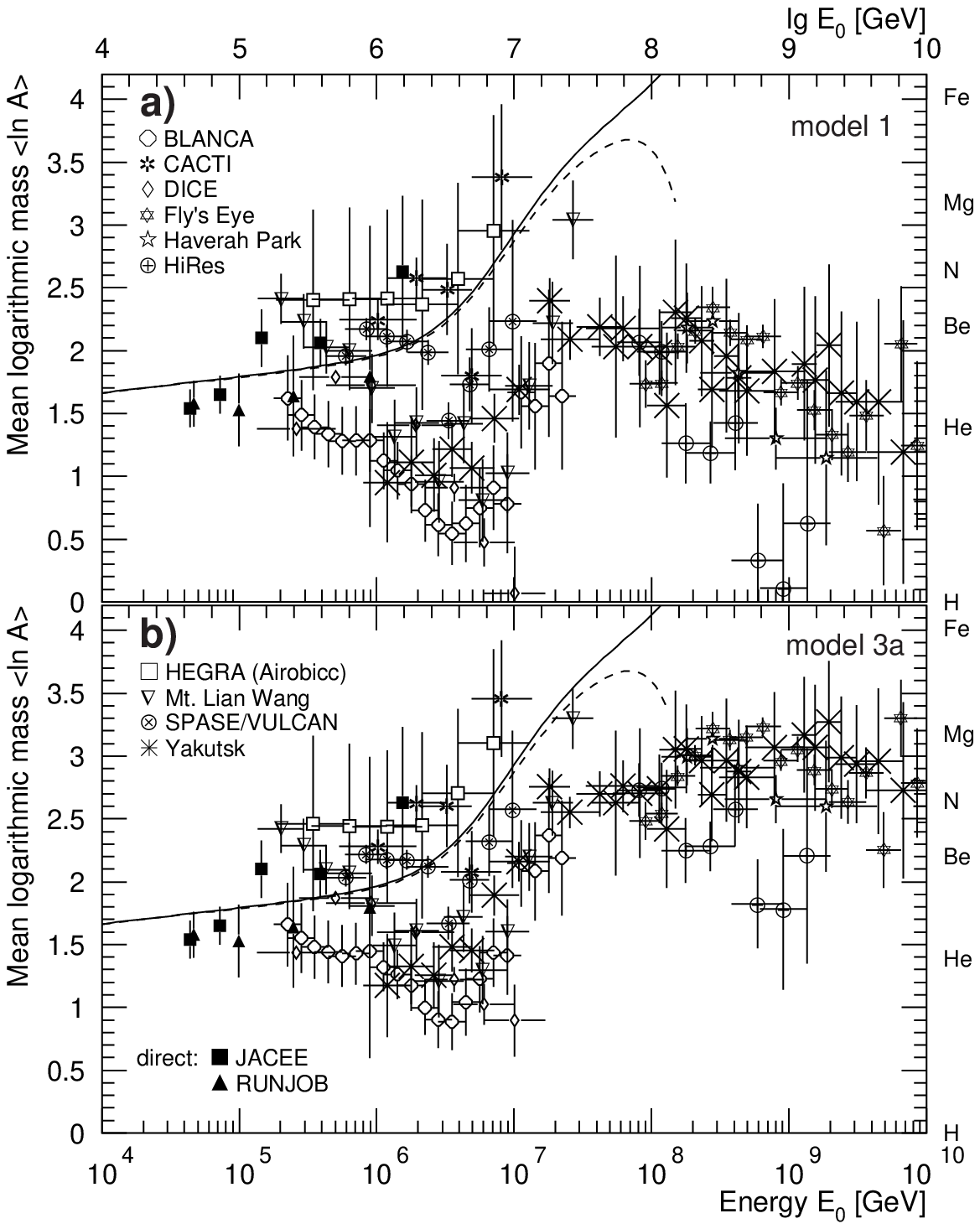}
\includegraphics{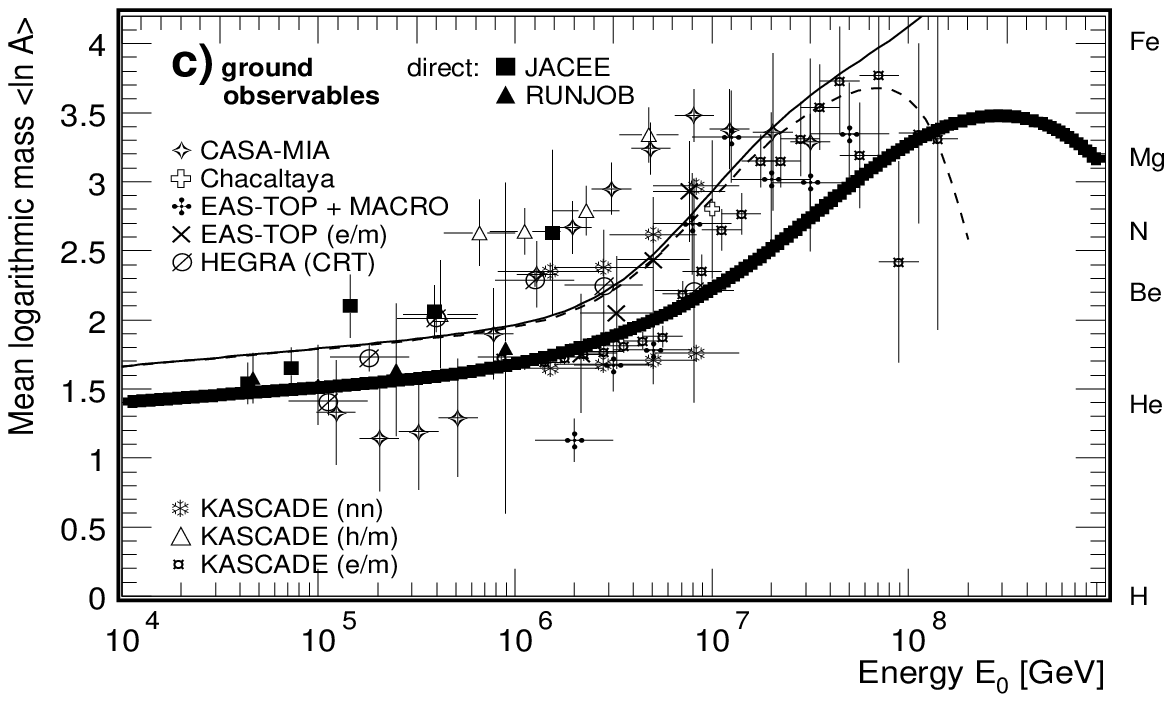}
\includegraphics{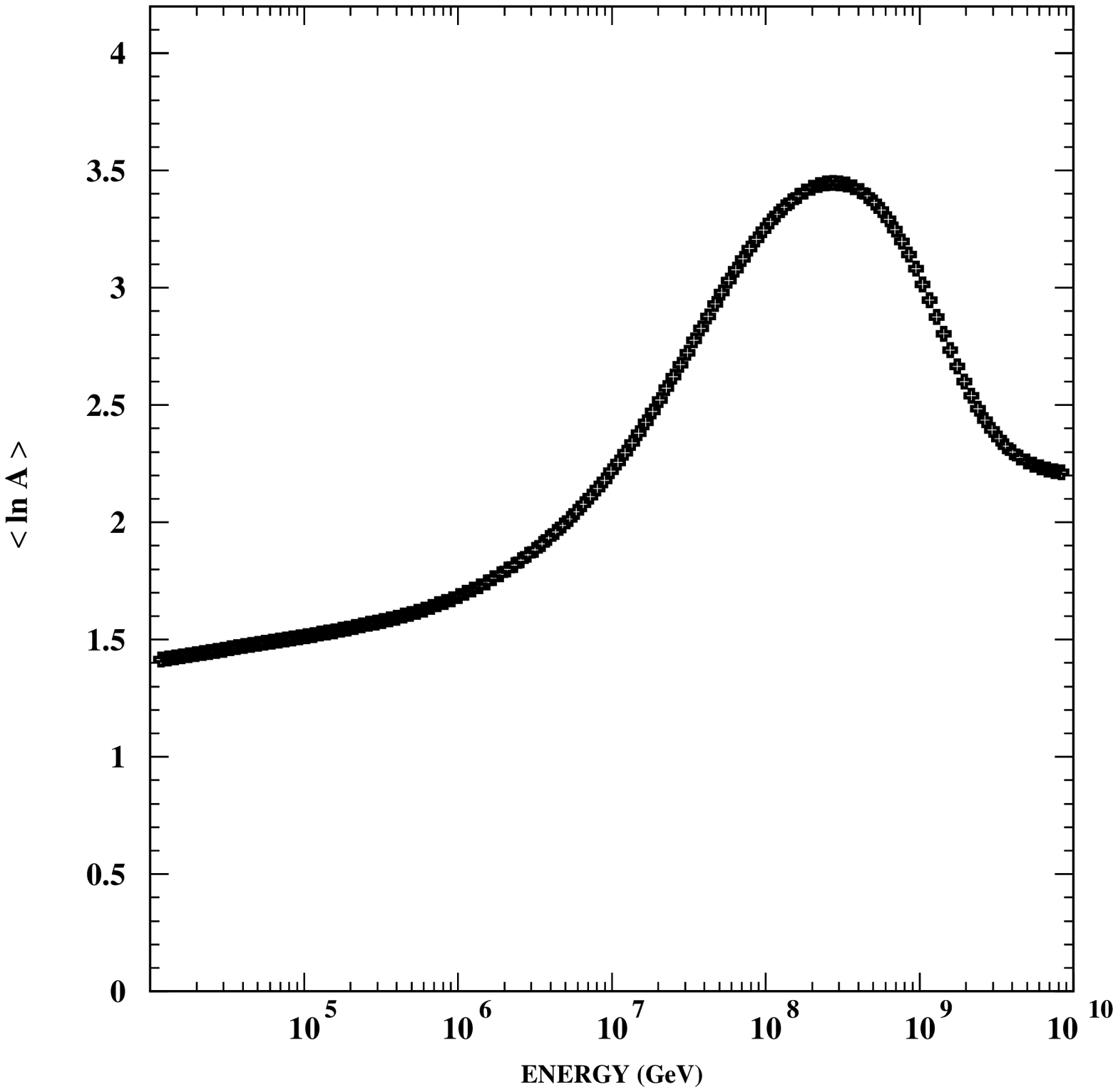}
\caption{Comparison between the CB model prediction for the mean 
logarithmic atomic mass $<lnA>$ as function of CR energy 
(thick lines of crosses on the rhs)
and $<lnA>$ as obtained by various experiments and compiled by Hoerandel 
2004: (a) from measurements of the average depth of 
shower maximum  interpreted with CORSIKA/QGSJET01, 
(b) interpreted with a modified
version, and (c) from  experiments measuring electrons, muons and 
hadrons at ground level.   
}
\label{lnA}
\end{figure}


{\footnotesize

\begin{thebibliography}{}

\bibitem{}
Bednarz, J. \&  Ostrowski, M. 1998, PRL, 80, 3991
\bibitem{} 
Biermann, P., \&  Sigl. G., 2001, Lect. Notes Phys. 576, 1
\bibitem{} 
Bird, D. J., et al. 1995, ApJ, 441, 144 
\bibitem{} 
Cronin, W. J. 2004, astro-ph/0402487
\bibitem{}
Dado, S.,  Dar, A. \& De R\'ujula, A. 2002, A\&A, 388, 1079
\bibitem{} 
Dar, A. 1998b, astro-ph/9809163
\bibitem{}
Dar, A. 1998a, Proc. XIIth Workshop on Perspectives in Particle Physics'
(ed. M. Greco) Aosta Valley, p. 23,  March 1-7, 1998, Italy
\bibitem{}
Dar, A. \& De R\'ujula, A. 2000, astro-ph/0008474
\bibitem{}
Dar, A. \& De R\'ujula, A. 2001a, MNRAS 323, 391
\bibitem{}
Dar, A. \& De R\'ujula, A. 2001b, ApJ, 547, L33  
\bibitem{}
Dar, A. \& De R\'ujula, A. 2003, astro-ph/0308248 
\bibitem{}
Dar, A. \& Plaga, R. 1999, A\&A, 349, 259
\bibitem{} 
Dermer, C. D. \&  Humi, M. ApJ. 2001, 556, 479
\bibitem{}
De R\'ujula, A. 1987,  Phys. Lett., 193, 514
\bibitem{} 
Greisen, K., 1966, PRL, 16, 748
\bibitem{} 
Grevesse, N. \& Sauval, A. J. 1998, Space Sci. Rev. 85, 161
\bibitem{} 
Heinz, S. \& Sunyaev, R. A. 2002, A\&A, 390, 751  
\bibitem{} 
Hoerandel, J. R., 2004, astro-ph/0402356 and these proceedings
\bibitem{} 
Kampert, K. H. et al. 2004, astro-ph/0405608 
\bibitem{} 
Kirk, J. G.,  Guthmann, A. W.,  Gallant, Y. A., Achtenberg, A.
2000, ApJ. 542, 235
\bibitem{} 
Lingenfelter, R. E., Higdon, J. C. \& Ramaty, R.  2000, AIPC, 528, 375
\bibitem{} 
Nagano, A.,  Watson, A. A., 2000, RMP, 72, 689  
\bibitem{} 
Milgrom, M. \& Usov, V. 1996, Astropart. Phys. 4, 365.
\bibitem{}
Mirabel, I. F. \& Rodr¡guez, L. F. 1999, ARA\&A, 37, 409
\bibitem{}
Olinto, A. V. 2004, astro-ph/0404114 
\bibitem{}
Plaga, R. 2002, NA, 7, 317 
\bibitem{}
Puget, J. L., Stecker, F. W. \& Bredekamp, J. H. 1976, ApJ, 205, 638
\bibitem{}
Rodr¡guez, L. F. \&  Mirabel, I. F. 1999, ApJ, 511, 398
\bibitem{}
Stecker, F. W. \&  Salamon, M. H. 1999, ApJ, 512, 521 
\bibitem{}
Strong, A. W. \& Mattox J. R. 1996, A\&A 308, L21 
\bibitem{}
Vietri, M. 1995, ApJ. 453, 883 
\bibitem{}
Waxman, E. 1995, PRL 75, 386
\bibitem{}
Wiebel-Sooth, B., Biermann, P. L. \& Meyer, H. 1998, A\&A, 330, 389
\bibitem{}
Watson, A. A. 2001, AIPC, 586, 817
\bibitem{}
Watson, A. A. 2003, in Astronomy, Cosmology and Fundamental Physics, 
Proceedings of the  ESO-CERN-ESA Symposium held in Garching, Germany, 4-7 
March 2002, p. 216.
\bibitem{} Zatsepin, G. T., Kuzmin, V.  A. 1966, JETP Lett.  4, 78
\end{thebibliography}
}
\end{document}